...

# Ultrawhite structural starch film for sustainable cooling†

Yang Liu,[a] Andrew Caratenuto,[a] Xuguang Zhang,[a] Ying Mu,[a] Youssef Jeyar,[b] Mauro Antezza [bc] and Yi Zheng [*ad]

Reducing human reliance on high-electricity-consuming cooling technologies like air conditioning is crucial for reshaping the global energy paradigm. Through utilizing natural starch gelatinization, freeze-drying and densification processes, we fabricated an ultrawhite cooling starch film with an ultrahigh solar reflectance of 0.96 and strong infrared emittance of 0.94. The porous structure of the cooling starch film, systematically controlled by the mechanical pressing processing, allows for effective scattering of solar radiation while emitting strongly during the atmospheric transparency window, thereby contributing to high-efficiency daytime radiative cooling capacity. Furthermore, the cooling starch film exhibits excellent mechanical tensile strength, measuring at up to 38.5 megapascals, which is more than twice the strength of natural wood. The ultrawhite radiative cooling starch film holds significant promise for optimizing cooling energy usage, especially in hot and arid climates.

## 1 Introduction

In recent years, unprecedented extreme heatwaves, attributed to global warming, have severely affected vast areas of the Earth, endangering the survival of human populations in extremely hot regions.[1,2] Consequently, the rapid escalation in electricity demand from compressor-based air conditioning systems has resulted in frequent grid overloads and environmental issues during the summer season,[3] thereby emphasizing the urgent need to explore alternative cooling methods that do not require electricity consumption. The energy-free passive daytime radiative cooling (PDRC) technique has become attractive for reducing building electricity consumption by reflecting solar radiation ($\lambda \sim 0.3$–$2.5$ μm) and dissipating heat from these PDRC structures and materials through the atmospheric transparency window ($\lambda \sim 8$–$13$ μm) into the ultracold outer space ($\sim 3$ K).[4–9] Recently, PDRC structures and materials with high solar reflectance $\bar{R}_{solar}$ and high infrared emissivity $\bar{\varepsilon}_{IR}$ have been widely investigated using bio-inspired structures,[10,11] precision-designed photonic structures,[12–14] micro/nanoparticle-based structures,[15,16] polymers[14,17,18] and pigmented paint film-based structures,[19–22] all of which have achieved subambient temperature even under strong sunlight. Among them, the design and fabrication methods for precise photonic PDRC structures and materials are often expensive or complex, posing a significant obstacle to their large-scale deployment for building purposes. Meanwhile, common polymers like polyvinylidene fluoride (PVDF), poly(methyl methacrylate) (PMMA), polydimethylsiloxane (PDMS), and polymethylpentene (PMP) are often chosen in various pore- and particle-based PDRC structures.[21,23–34] However, these materials may release micro- and nanoplastics into the environment as they photodegrade after prolonged UV exposure. Additionally, even wood-based PDRC materials face challenges in large-scale production due to the lengthy wood regeneration process.[10] Therefore, identifying natural materials with shorter regeneration cycles for fabricating PDRC materials is crucial to advance the widespread implementation of environmentally friendly and sustainable radiative cooling technology.

Starch stands out as one of nature's most abundant natural polymers. Its widespread use in the food industry is attributed to its affordability, abundance, biodegradability, and edibility.[35] Structurally, starch comprises amylose and amylopectin. Amylose contains α-1,4 glycosidic bonds, while amylopectin contains both α-1,6 and α-1,4 glycosidic bonds.[36] Significantly, owing to the exceptional film-forming properties of amylose, cost effective and biodegradable starch films are widely used as packaging materials in the food industry.[37] In response to the escalating focus on addressing plastic waste associated with plastic-based PRDC materials and the growing demand for cost-effective and eco-friendly alternatives, as well as the urgent need for large-scale production, we have employed natural potato starch with high amylose content in the fabrication of PRDC materials through a series of processes including gelatinization, liquid nitrogen ($N_2$) freezing, freeze-drying, and mechanical pressing. The ultrawhite cooling starch film obtained, featuring abundant pores as shown in Fig. 1, demonstrates outstanding solar reflectance (0.96) and remarkable infrared emittance (0.94). This translates to exceptional passive radiative cooling performance, showcasing average and peak outdoor temperature reductions of 6.8 °C and 11.3 °C, respectively. Moreover, the cooling starch film sustains an outstanding radiative cooling power of 87 W m$^{-2}$ in continuous daytime cooling tests. In addition, the gelatinized and densified cooling starch film exhibits mechanical strength that is more than twice as strong as that of natural wood. With these advantages, the porous cooling starch film will emerge as a highly promising renewable material candidate for energy-efficient, environmentally friendly and sustainable passive radiative cooling technology.

[a]Department of Mechanical and Industrial Engineering, Northeastern University, Boston, MA 02115, USA. E-mail: y.zheng@northeastern.edu
[b]Laboratoire Charles Coulomb (L2C), UMR 5221 CNRS-Université de Montpellier, F-34095 Montpellier, France
[c]Institut Universitaire de France, 1 rue Descartes, F-75231, Paris Cedex 05, France
[d]Department of Chemical Engineering, Northeastern University, Boston, MA 02115, USA

† Electronic supplementary information (ESI) available. See DOI: https://doi.org/10.1039/d4ta08793c

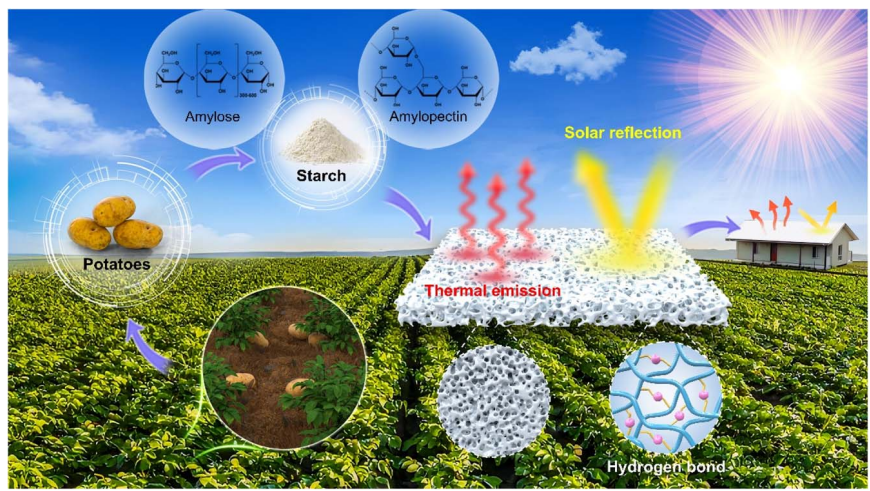

Fig. 1 Concept of the natural starch-derived sustainable radiative cooling film and its high-efficient radiative cooling mechanism. Cooling starch film with high $\bar{R}_{\text{solar}}$ and $\bar{\varepsilon}_{\text{IR}}$ enabling subambient cooling under direct sunlight. Abundant structural micro- and nanopores contribute to effective sunlight backscattering, while the inherent molecular vibrations of chemical bonds in starch result in strong infrared thermal emission within the atmospheric transparency window.

## 2 Results and discussion

### 2.1 Material fabrication and characterization

Fig. 2 depicts the fabrication process of cooling starch film. Initially, potato starch was dissolved in hot deionized water while being mechanically stirred until a homogeneous and translucent starch paste solution formed (Fig. S1†).

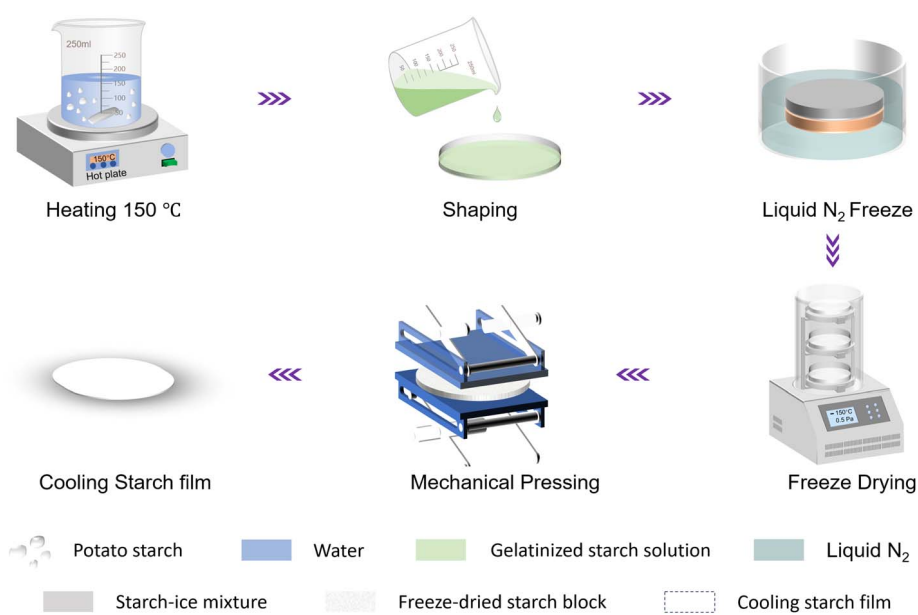

Fig. 2 Schematic illustrating the fabrication process of the cooling starch film. The fabrication process includes starch gelatinization, shaping, liquid $N_2$ freezing, freeze-drying, and mechanical pressing processes.

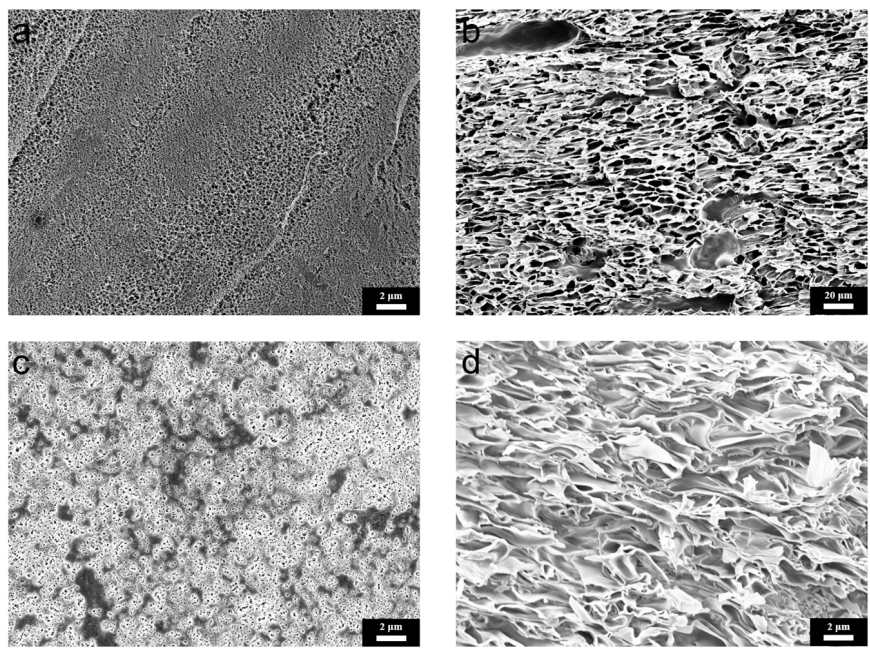

Fig. 3 Surface characterizations of cooling starch film. SEM images of (a) bottom and (b) cross-sectional view of starch film before pressing, and (c) bottom and (d) cross-sectional view of starch film after pressing.

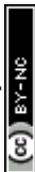

Subsequently, the gelatinized starch solution was frozen into a starch-ice block in a mold, followed by freeze-drying in a freeze dryer to form an ultrawhite porous starch block. Finally, the porous white block was mechanically pressed into a hard cooling starch film. Details of the material and method are provided in ESI S1.† We utilized scanning electron microscopy (SEM) to visualize the porous structures of before- and after-pressing cooling starch films, as seen in Fig. 3. These surface and internal pores serve as randomized and disordered scattering elements for incident sunlight, resulting in a strong, wide-ranging solar reflection. Although the mechanical pressing can fuse or eliminate some surface pores and decrease the surface porosity of cooling starch films (Fig. 3a and c), this process can reduce the sizes and then increase the density of the internal pores in the cooling starch film, which can be clearly seen from the cross-section views of before- and after-pressing starch films (Fig. 3b and d), thereby enhancing the solar reflection within visible wavelength range, as shown in Fig. 4a. Our solar reflectance analysis of the cooling starch films, taken at an incident angle of 6°, reveals an exceptionally high average solar reflection of 96% for densified cooling starch film, highlighting its strong scattering properties. The high solar reflectance is responsible for the bright whiteness observed in the cooling starch film (Fig. 7a). We also investigated the thermal emissivity of the cooling starch film within the infrared range, specifically between 2.5 to 18 μm, which encompasses wavelengths crucial for room-temperature blackbody radiation. The cooling starch film exhibits ultrahigh infrared emissivity (~0.94), emitting infrared radiation strongly through the atmospheric transparency window (8–13 μm) into the cold outer space. Consequently, the cooling starch film appears black in the infrared spectrum, a distinct contrast from its white appearance under sunlight exposure, indicating its significant potential for passive radiative cooling. Furthermore, Fourier transform infrared (FTIR) transmission spectra demonstrate that the cooling starch film exhibits strong emission from 8 to 13 μm (Fig. 4b and c), primarily assigned to OH association and stretching vibrations of C–H, C–O, and C–O–C bonds of starch coinciding with the atmospheric transparency window.[38,39]

To investigate the effects of surface morphology and porous structure of cooling starch films on solar reflection and infrared thermal emission, we further analyzed the spectral characteristics of cooling starch films fabricated under varying mechanical pressures (0 MPa, 30 MPa, 40 MPa, and 50 MPa, respectively), as shown in Fig. 5. The optical properties of samples under other pressure conditions (10 MPa and 20 MPa) are presented in Fig. S2†. Fig. 5a demonstrates their solar reflectance of four cooling starch films (Starch$_{0MPa}$, Starch$_{30MPa}$, Starch$_{40MPa}$, and Starch$_{50MPa}$, respectively). Obviously, not all compressed starch films have higher solar reflectance than the uncompressed starch sample (Starch$_{0MPa}$, with the solar reflectance $\bar{R}_{solar}$ of 0.94). The Starch$_{30MPa}$ exhibits the highest average solar reflectance, with $\bar{R}_{solar}$ of 0.96. In contrast, the solar reflectance of Starch$_{40MPa}$ and Starch$_{50MPa}$ subjected to higher pressures decreases as the mechanical pressure increases, to 0.94 and 0.92, respectively. This is because the mechanical pressing with proper pressure (e.g., 30 MPa) can increase the density of the internal pores in the cooling starch film (Fig. 3b and d), which can offset the disadvantage caused by decreased surface pores due to the pressing, thereby enhancing the reflection of sunlight, as shown in Fig. 5a. However, excessively high pressures (e.g., 40 MPa and 50 MPa) will destroy more surface pores and significantly decrease the surface porosity of cooling starch films, causing the originally porous cooling starch films tend to be a planar structure. This will reduce the pore-driven scattering of sunlight, as shown in Fig. 6. Furthermore, Fig. 5b shows that these starch films exhibit high infrared emissivity during 8–13 μm. On the contrary, the Starch$_{0MPa}$ presents the highest average thermal emittance $\bar{\varepsilon}_{IR}$ = 0.95, while other pressed starch films drop gradually to lower average thermal emittance in the same wavelength range of 0.94, 0.93, and 0.93 for Starch$_{30MPa}$, Starch$_{40MPa}$, and Starch$_{50MPa}$, respectively. This phenomenon can be attributed to the abundant micro- and nanopores in Starch$_{0MPa}$, which facilitate smoother gradient transitions in the refractive index at the starch–air interface. This results in reduced surface reflectance and enhanced thermal emissivity for the unpressed Starch$_{0MPa}$ in the range of 8–13 μm.[41,42] In contrast, mechanically pressed starch films with fewer widely distributed micro- and nanopores exhibit sharper transitions in the refractive index across the interface, as seen in Fig. 3b, d and 6. Consequently, the Starch$_{30MPa}$ exhibits enhanced surface reflectance and slightly reduced thermal emissivity in mid-IR wavelengths due to the application of proper external mechanical pressure. Table S1† presents their density and porosity, highlighting the differences among Starch$_{0MPa}$, Starch$_{30MPa}$, Starch$_{40MPa}$, and Starch$_{50MPa}$. Therefore, the porous structure with appropriate pore size and porosity can maintain a high-performance balance between high solar reflection and high infrared emission, thereby effectively enhancing the radiative cooling limit of existing porous radiative cooling structures.

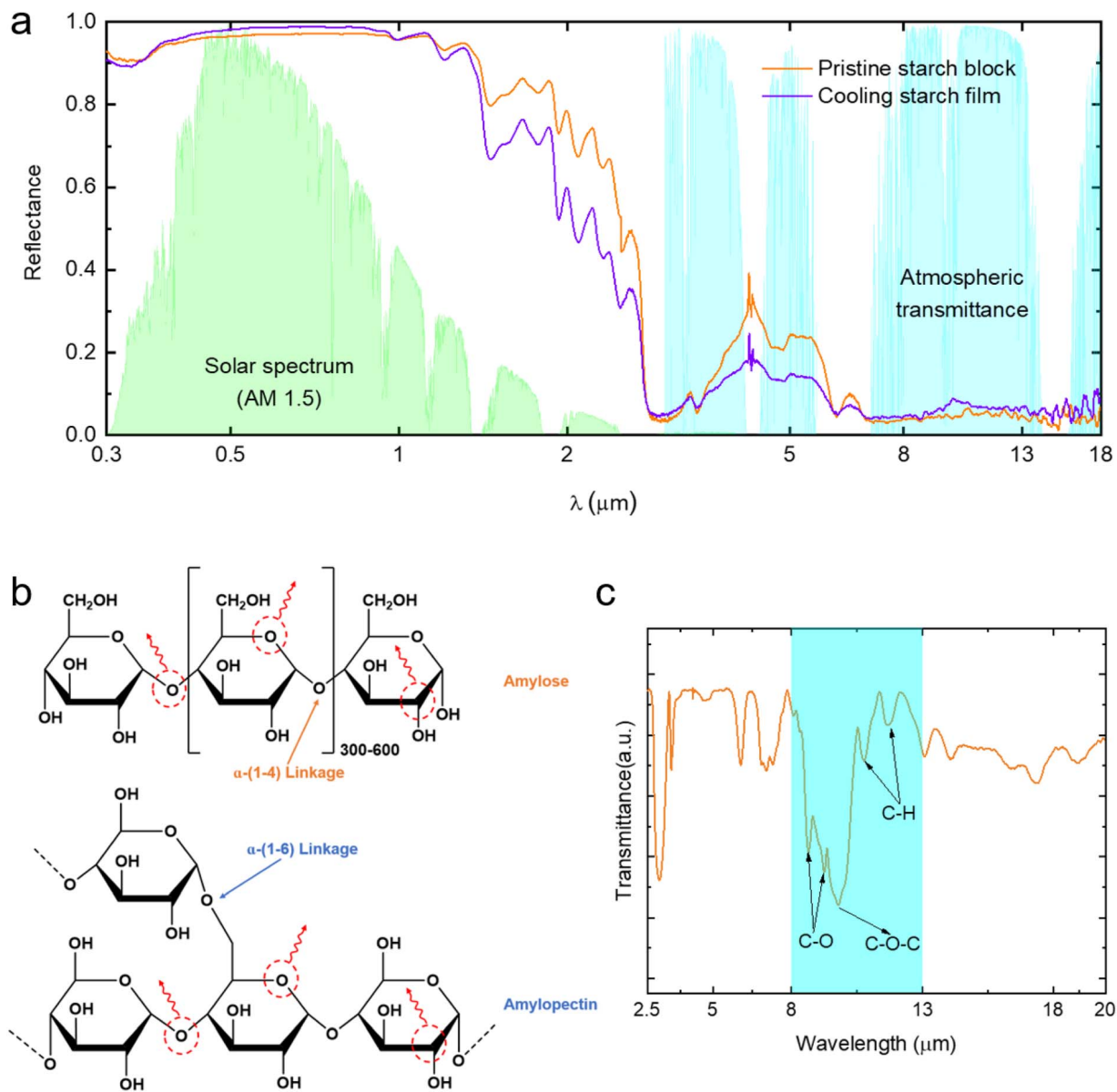

Fig. 4 Radiative cooling mechanism of cooling starch film. (a) Spectral reflectance of pristine starch block with a thickness of 8 mm and pressed cooling starch film with a thickness of 0.8 mm. (b) Schematic showing the molecular structures of amylose and amylopectin in the starch.[40] (c) FTIR transmittance spectrum of the starch presenting OH association and stretching vibrations of C–H, C–O, and C–O–C absorption bands.

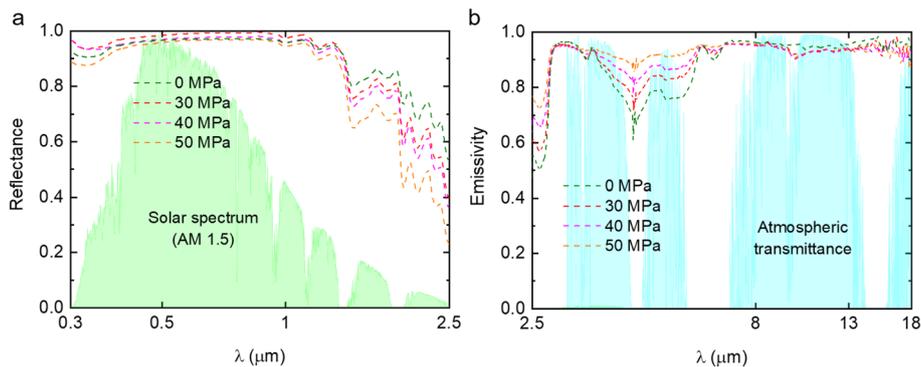

**Fig. 5** Optical properties of four cooling starch films under different mechanical pressures. (a) Solar reflectance and (b) infrared emissivity of Starch$_{0MPa}$, Starch$_{30MPa}$, Starch$_{40MPa}$, Starch$_{50MPa}$, respectively.

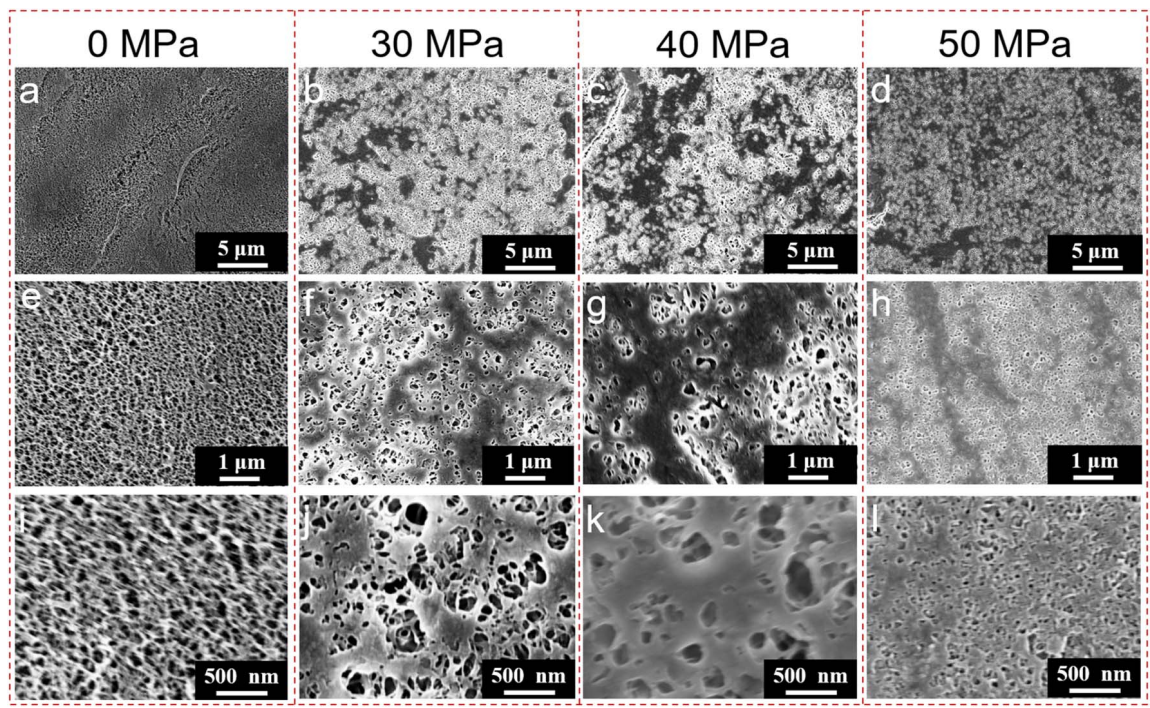

**Fig. 6** Surface characterizations of unpressed and pressed cooling starch films. SEM images of bottom view of (a, e and i) Starch$_{0MPa}$, (b, f and j) Starch$_{30MPa}$, (c, g and k) Starch$_{40MPa}$, and (d, h and l) Starch$_{50MPa}$ at different magnifications, respectively.

### 2.2 Subambient radiative cooling performance of cooling starch film

To evaluate the radiative cooling capability of the cooling starch film, we conducted outdoor cooling experiments. Specifically, a cooling starch film (Starch$_{30MPa}$) with a 65 mm diameter and 0.8 mm thickness was continuously monitored from 9:00 AM to 5:00 PM on September 04, 2023, on a 30 meter-high building at Northeastern University in Boston, Massachusetts (coordinates: 42.36° N, 71.06° W), as depicted in Fig. 7. Two balsa wood samples (one with commercial white paint and one without) were used as comparison groups during the outdoor cooling test. The inset of Fig. 7a compares the spectral reflectance of these three samples. A commercial ambient weather station was employed to monitor solar intensity (~590 W m$^{-2}$), wind speed (~1.2 m s$^{-1}$), and relative humidity (~48%), respectively, as shown in Fig. 7a–c. After analysis of the experimental data in Fig. 7d, it is found that the cooling starch film achieved an average sub-ambient cooling performance of 6.8 °C over the 8 hours measurement period and demonstrated a maximum temperature decrease of 11.3 °C below the ambient under a solar intensity of 643 W m$^{-2}$. The cooling starch film also achieved an average subambient cooling of 2.8 °C at night (Fig. S3†). Meanwhile, the average temperatures of the bare balsa wood sheet and the painted wood sheet were 6.4 °C and 4.7 °C higher, respectively, than that of the cooling starch film, thereby highlighting the outstanding radiative cooling capacity

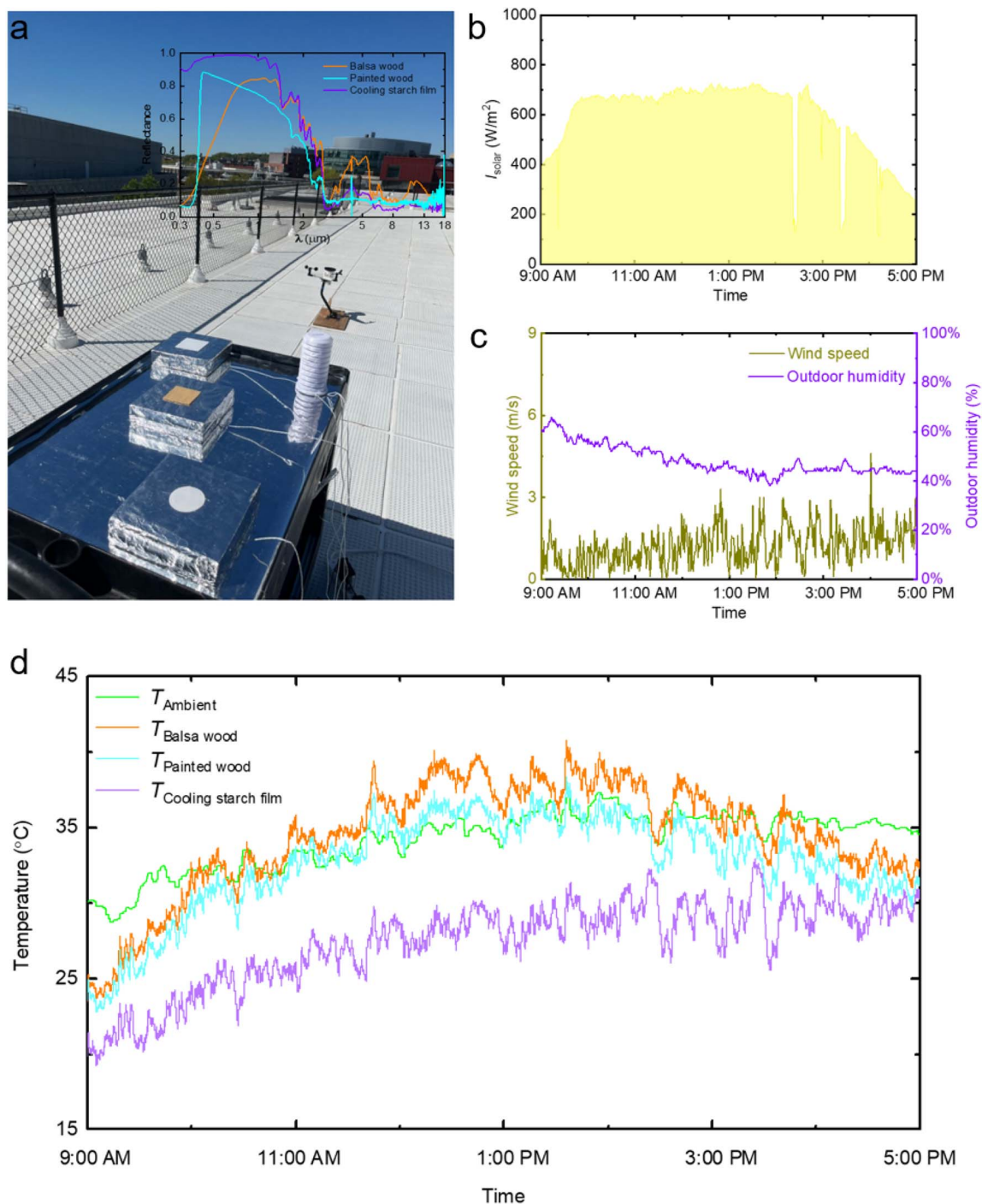

Fig. 7 Radiative cooling experiment of the cooling starch film. (a) Real-time temperature measurement setup, Boston, USA (42.36° N, 71.06° W) on September 04, 2023. (b and c) Real-time Solar intensity, wind speed, and relative humidity. (d) Temperatures of different samples (wood sheet, painted wood sheet, and cooling starch film). The inset of (a) shows the spectral reflectance of wood sheet, painted wood sheet, and cooling starch film, respectively.

of the cooling starch film when contrasted with typical building materials. The radiative cooling performance of the cooling starch film is comparable to the previously reported PDRC films (Table S2†). Considering the outdoor applications of cooling starch film, several well-established methods can improve its waterproofing, such as encapsulation with transparent commercial films (e.g., polyethylene thin films),[43] polymeric coatings (e.g., air-spraying plastic microparticles atop the materials),[23] and chemical surface treatment (e.g., fluorosilane treatment).[10] Firstly, we wrapped the cooling starch film with a commercial PE film to enhance its waterproofing. Fig. S7† presents the spectral reflectance of the cooling starch film with and without a commercial PE film covering. While the commercial PE film slightly reduces infrared emission, it does not significantly affect the solar reflectance of the cooling starch film. Additionally, we also presented an additional method to enhance the waterproofing of the cooling starch film by air-spraying ethanolic poly(tetrafluoroethylene) (PTFE) microparticle suspension on the cooling starch film. The PTFE coating can significantly improve the hydrophobicity of the cooling starch film. For detailed experimental methods refer to the literature.[23] Although the PTFE coating slightly reduces the solar reflectance of cooling starch film, it transforms the cooling starch film from a hydrophilic to a highly hydrophobic (water contact angle $\theta = 147°$) (Fig. S8 and Video S1†). This feature enables the cooling starch film to have a waterproof surface, making it suitable for outdoor applications. Therefore, these methods potentially address the waterproofing challenge for the outdoor application of the cooling starch film. More discussion on the stability of the cooling starch film for outdoor application is provided in ESI S3.†

While these strategies can enhance the durability of the cooling starch film, they often involve costly or environmentally harmful synthetic chemicals and plastic components, which contradict the core motivation of this work. Our primary objective is to highlight the intrinsic optical properties and radiative cooling capacity of pure natural starch film. Therefore, we avoided introducing additional materials for durability enhancement in this study. This work represents an initial step in transforming a single natural material into a radiative cooling material, paving the way for sustainable cooling solutions. In the future, we will explore three key directions to improve the long-term outdoor applicability of cooling starch films: (1) applying hydrophobic and UV-resistant coatings using natural inorganic materials, (2) modifying the chemical structure of starch and optimizing the structure of starch-based radiative cooling films to ensure sustained outdoor performance, and (3) expanding the application scope of starch-based materials by leveraging their excellent optical and mechanical properties.

Furthermore, we calculated the cooling power of the cooling starch film during outdoor cooling testing, which considers the measured temperatures of the cooling starch and ambient, as well as outdoor weather conditions. Details of the calculation are provided in ESI S2.† The cooling starch film demonstrated the radiative cooling power around 87 W m$^{-2}$ from 9:00 AM to 5:00 PM on September 04, 2023, in Boston (Fig. 8a), displaying the outstanding radiative cooling performance achieved through the gelatinization/freeze-drying/pressing process of starch. Additionally, Fig. 8b and c depicts the net cooling power ($P_{cool}$) of cooling starch film during daytime and nighttime, respectively. These results are presented as functions of the cooling starch film's temperature under various non-radiative heat coefficients ($h_{nr}$ = 0, 2, 4, 6, 8, and 10 W m$^{-2}$ K$^{-1}$),[41,44] with the ambient temperature $T_{Ambient}$ set at 25 °C. These two figures clearly illustrate that as the temperature of the cooling starch film increases, the cooling power progressively rises until reaching a balance with ambient temperature. During daytime conditions, the maximum net cooling power can reach 77.8 W m$^{-2}$, while nighttime cooling power can reach up to 119 W m$^{-2}$. The lower value of the daytime cooling power is attributed to sunlight absorption partially offsetting the cooling starch film's cooling capacity. Additionally, when the non-radiative heat coefficient is 10 W m$^{-2}$ K$^{-1}$, the cooling starch film temperatures decrease to 5.3 °C (daytime) and 8.2 °C (nighttime), respectively. This highlights the significant effect of heat transfer from the ambient environment on real-life cooling performance. Both the theoretical and experimental results showcase the potential of the cooling starch film for highly efficient radiative cooling applications.

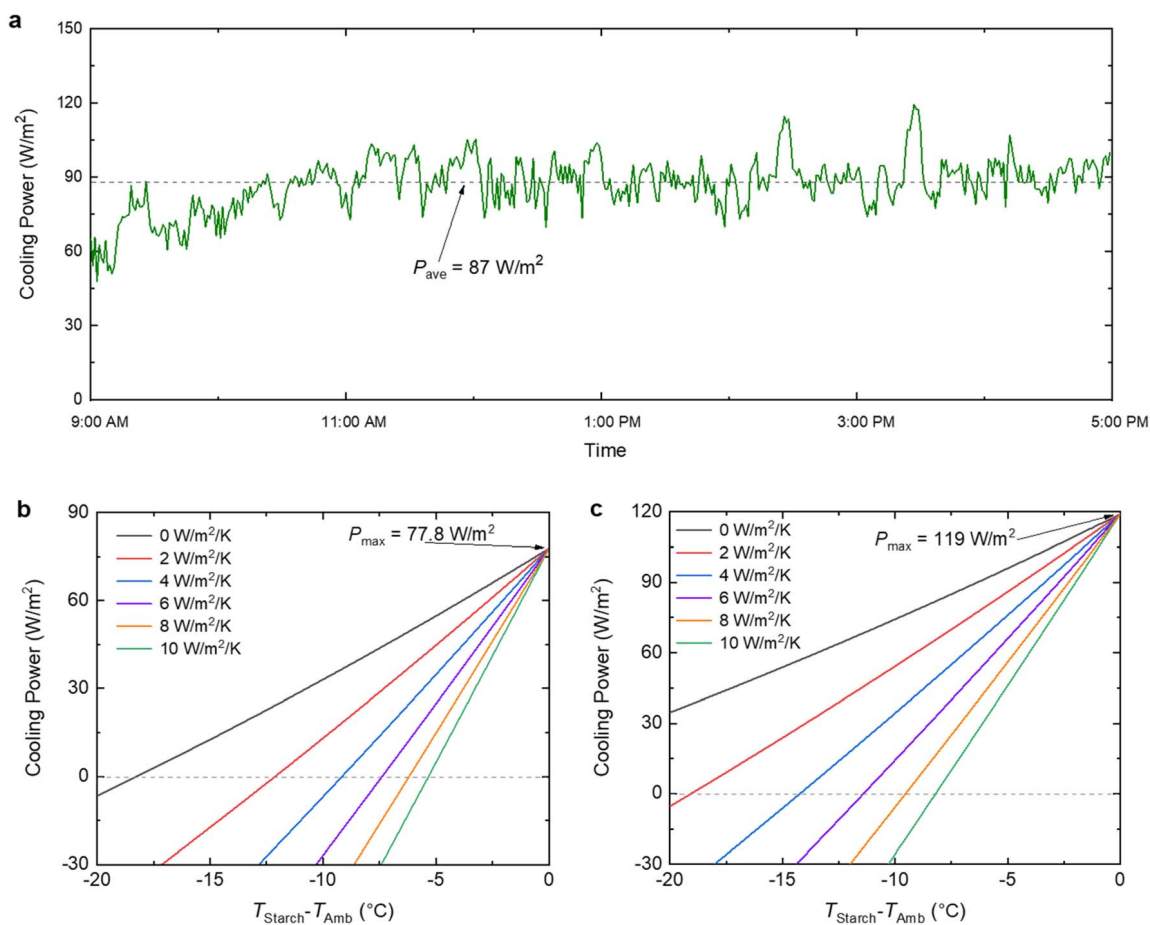

Fig. 8 Theoretical calculation for cooling power. (a) Cooling power of the cooling starch film on September 04, 2023, in Boston. Calculated net (b) daytime and (c) nighttime cooling power of cooling starch film, respectively, where $h_{nr}$ = 0, 2, 4, 6, 8 and 10 W m$^{-2}$ K$^{-1}$.

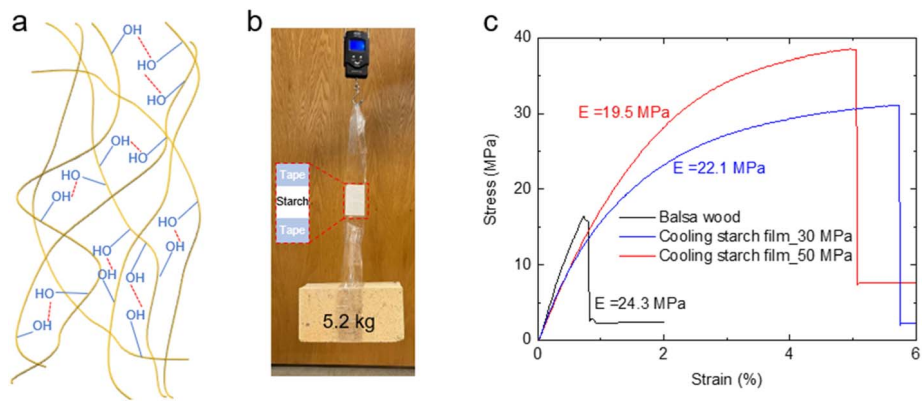

Fig. 9 Superhard cooling starch. (a) Schematics showing the high mechanical strength driven by the intermolecular hydrogen bonds of the starch. (b) A free-standing cooling starch film with 50 mm × 30 mm × 0.8 mm can easily bear a load of 5.2 kg. (c) Stress–strain curves of 0.8 mm thick piece of the cooling starch film compared with natural wood.

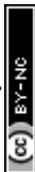

### 2.3 Mechanical properties of the cooling starch film

During the gelatinization process, potato starch powders absorb water and swell. The hydrogen bonds between starch molecules break, leading to the formation of a uniform colloidal solution. When the gelatinized starch is dried, the water gradually evaporates, and the amylose and amylopectin molecules recombine. In particular, the relatively high content of amylose in potato starch forms a tight and strong network structure through hydrogen bonds. As the starch film dries, intermolecular hydrogen bonding between starch molecules becomes more pronounced, further enhancing the mechanical strength of the starch film. Besides, the porous structural shrinkage during freeze-drying, combined with the subsequent external force pressing into the film, reduces the pores and voids in the cooling starch film, thereby making it denser and harder. The cooling starch film exhibits greater mechanical strength compared to natural wood due to the reformation of hydrogen bonds of the starch after starch gelatinization, as illustrated in Fig. 9a, S4 and S5.† Meanwhile, a free-standing cooling starch film (Starch$_{30MPa}$) with 50 mm × 30 mm × 0.8 mm can effortlessly bear a load of 5.2 kg, demonstrating its potential for use in engineering applications (Fig. 9b). Furthermore, the mechanical pressing of cooling starch increases the interaction area of hydroxyl bonds, thereby further enhancing its mechanical strength. Therefore, the cooling starch film (Starch$_{50MPa}$) demonstrates a tensile strength as high as 38.5 MPa, which is ∼2.34 times that of wood sheet of the same thickness (Fig. 9c). Therefore, the cooling starch film outperforms natural wood in building thermal applications due to its excellent cooling capability and exceptional mechanical strength, making it an appealing option as a structural cooling material when compared to other radiative cooling materials.

## 3 Conclusions

In conclusion, we have demonstrated a pure-natural starch-based passive radiative cooling material fabricated by the gelatinization/freeze-drying/pressing method to tailor its spectral response. This cooling starch film showcases exceptional whiteness, derived from its disordered porous structure that can be systematically controlled by mechanical pressing. The emitted energy within the infrared range from the cooling starch film surpasses solar and environmental energy gains, thereby contributing to excellent daytime radiative cooling performance. We have validated this subambient cooling effect through outdoor measurements of cooling starch films. Moreover, the cooling starch film demonstrates remarkable mechanical properties, being 2.34 times stronger than wood sheet of the same thickness. Its potential impact includes substantial reductions in micro- and nanoplastic emissions and cooling electricity consumption, thereby contributing significantly to environmentally conscious and sustainable cooling technology.

## Data availability

The data that support the findings of this study are available within the article and its ESI file.† All other relevant data supporting the findings of this study are available from the corresponding authors upon request.

## Author contributions

Y. L., M. A., and Y. Z. conceived the initial concept. Y. L., A. C. X. Z. performed sample fabrication and radiative cooling experiments. Y. L., Y. M., and Y. J. conducted materials characterization experiments. All authors discussed and analyzed the data and results. Y. L. wrote the original draft. Y. Z. supervised this project.

## Conflicts of interest

There are no conflicts to declare.

## Acknowledgements

This project is supported by the National Science Foundation through grant number CBET-1941743.

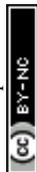